# A comparison of cluster algorithms as applied to unsupervised surveys

Kathleen Campbell Garwood*

Saint Joseph's University,
5600 City Ave, Philadelphia, PA 19131, USA
Email: kcampbel@sju.edu
*Corresponding author

Arpit Arun Dhobale

Indian Institute of Technology,
Near Doul Gobinda Road, Amingaon,
North Guwahati, Guwahati, Assam – 781039, India
Email: appu.dhobale@gmail.com
Email: dhobale.arun@alumni.iitg.ac.in

**Abstract:** When considering answering important questions with data, unsupervised data offers extensive insight opportunity and unique challenges. This study considers student survey data with a specific goal of clustering students into like groups with underlying concept of identifying different poverty levels. Fuzzy logic is considered during the data cleaning and organising phase helping to create a logical dependent variable for analysis comparison. Using multiple data reduction techniques, the survey was reduced and cleaned. Finally, multiple clustering techniques (k-means, k-modes and hierarchical clustering) are applied and compared. Though each method has strengths, the goal was to identify which was most viable when applied to survey data and specifically when trying to identify the most impoverished students.

**Biographical notes:** Kathleen Campbell Garwood is an Assistant Professor in the Department of Decision and System Sciences and has been a member of the faculty at the SJU since 2004. Her teaching primarily focuses in data mining and modelling with the goal of introducing real data and analytical techniques to students. Her research interests include data visualisation (best practices and techniques), rank order comparisons (with a focus on sustainability rankings), modeling applications with real world settings (including identifying the most impoverished for the Fe y Alegria: Bolivia) and data collection and analysis related to science, technology, engineering and math (STEM). Recently, she has combined these interests working with the United Nations' principles for responsible management education (PRME) group to collect organise and visualise through interactive dashboards the efforts being made in management school to address the UN sustainable development goals. In this venture, she works to oversee and educate undergraduate and graduate students in data collection and visualisation.

Arpit Arun Dhobale is a graduate from the Indian Institute of Technology Guwahati. He is currently working at the EXL Services as part of the Data Science team. He enjoys working in the field of data analytics and is enthusiastic about growing his skills in predictive analytics with a goal of accuracy and implementable results. While in school, he worked on an interesting project on improving the hotel utilising rate using data analytics. He is fascinated by the work currently done in the field of artificial intelligence in the present world. Having hand on experience in various machine learning techniques, he has gained hegemony in this particular area.

## 1 Introduction

Often survey analysis collects data to try to identify response patterns leading to groupings of respondents with different characteristics as revealed by answers provided to survey questions. Without additional background information on respondents, it is often very difficult (and many times impossible) to verify the accuracy of groupings resulting from the analysis. This paper examines one such situation in which high school students in low-income neighbourhood schools in Bolivia responded to a standard periodic institutional survey and responses were analysed to better understand respondents' socio-economic contexts. In this case study, the question to be answered was "can we identify the most impoverished students based on a 22 questions standard survey alone?". With no known dependent variable and an inability to objectively capture the socio-economic condition of the students being surveyed, the task of coming to a conclusive answer becomes unfeasible as there is no way to validate at least some portion of the students identified as most impoverished. In order to assess the accuracy of accepted statistical practices of clustering to identify class (Vyas and Kumaranayake, 2006), a 'baseline' for poverty using UN definitions was established. After the World Summit of Social Development in Copenhagen in 1995, 117 countries adopted a declaration and programme of action which included committees to eradicate 'absolute' and reduce 'overall' poverty. Absolute poverty was defined as: 'a condition characterised by severe deprivation of basic human needs, including food, safe drinking water, sanitation facilities, health, shelter, education and information [UN, (1995, p.57]. Using the distributions of each question pertaining directly to this United Nations definition of absolute poverty, those students who answered in the lowest 5 percentiles were indicated as impoverished, therefore establishing a baseline dichotomous dependent variable through which lowest threshold students were indicated as a one (extreme poverty) and all others as a zero.

Using the survey questions directly related to food (number of meals a day) and shelter (number of rooms in the household) while also considering access to water and electricity, distributions of the data were made and fuzzy logic was applied to single out the students who fell within the lowest 5% as shown herein in Algorithm 1. This survey is an instrument used by a network of schools in low income neighbourhoods in different cities in Bolivia, hence though all students could be classified as impoverished, the analysis sought to identify those in most need with two goals. First, the objective was to identify the most impoverished students in each cohort to better target additional outreach support. Second, changes in the distributions were expected depending on the location (i.e., region within Bolivia) as different contextual conditions might imply different indicators of extreme poverty because of different cultural, climactic, historical conditions and ideally the logic system should allow thresholds to adjust with the data. Moreover, the overarching goal of this data collection practice is to find the most impoverished group including but NOT limited to the students who fall in the range of absolute poverty. While these students are of interest to establish at least a reasonable dependent variable, the cluster techniques applied to the data are meant to capture more than just these baseline students.

## 2 Literature review

When considering approaches to identify a solution or group within a survey without access to information allowing result verification, exploratory factor analysis (EFA) has become a useful and accepted approach. This methodology helps narrow down the important questions that provide the most variability among those taking the survey. In the social sciences this technique has found a variety of applications where it beneficially detects the proper variables to be considered as the data is being reduced. Recent studies include constructing socio-economic status indices (Vyas and Kumaranayake, 2006); a methodology for evaluating school principals (Lovett et al., 2002); assessing variables that might motivate high-school students (Morris, 2001); and helping determine services which would provide college students the best experience (Major and Sedlacek, 2001). Osborne and Costello (2005) outline 'best practices' in exploratory factor analysis for reducing the number of questions considered from the survey as well as the ordering for the next phase of the analysis.

Though EFA is a good first step for narrowing the data – it does not completely address the needed result of forming groups of like respondents, which is the goal of this research. Using the factors indicated from the exploratory analysis, the next relevant statistical methodology is cluster analysis. This technique allows for the partitioning of objects (in this case respondents) in such a way as to minimise variation between objects placed into any given group (called a cluster) while making sure they are sufficiently dissimilar to objects classified in another group (or cluster). Though clusters are often not clearly separated from one another, most cluster analyses aim for a crisp classification of the data into non-overlapping groups (Sharma et al., 2012).

Human logical reasoning, when applied to data, allows for inferences and interpretations of the results but take time when in some cases automation is feasible. However, when these inferences can change slightly over time or with changing data, a method of accounting for changes and adapting the results is the motivation behind the choice of using fuzzy logic for this paper. Fuzzy logic as applied to this data is a method to partition the respondents by applying a set of rules which can change depending on the data source and responses given as the survey evolves. Recent literature (Lughofer et al., 2017) suggests that with gradual shifts and gradual growth (in this case of the survey questions and places where responses are collected) that the rules may grow larger over time based on new incoming samples. This process of growth will take time and vary as other countries are incorporated in the study and the surveys are expanded, it will be important to consider how to transition the partitioning. Moreover, fuzzy logic as applied to unsupervised data requires training and human intervention though recent strides have been made to help automate this process and allow for steps to be made for automatic clustering or in this case partitioning of the data (AL-Sharuee et al., 2018). This application of fuzzy logic was used to create a usable dependent variable and techniques to adjust it with the data may be considered in future work.

Cluster analysis is an important data-mining technique used to segment data and find patterns (Wei-ning and Ao-ying, 2002). Their formal definition is as follows:

*Definition 1:* Given a data set $V\{v_1, v_2, \dots, v_n\}$, in which $v_i$'s$(i = 1, 2, \dots, n)$ represent each data point. The process of partitioning $V$ into $\{C_1, C_2, \dots, C_k\}$, $C_i \subseteq V$ $(i = 1, 2, \dots, k)$ and $\cup_{i=1}^{k} C_i = V$, based on similarity between data points are clustering, $C_i$'s$(i = 1, 2, \dots, k)$ are called clusters.

The way that similarities among data points are grouped is not well defined which leads to different methods each following different criteria. Wei-ning and Ao-ying (2002) propose attempting several methods in order to assess which is the most useful for the data at hand. This unsupervised learning process allows for analysis without a priori knowledge of the data set, where the quality of the result is important (Wei-ning and Ao-ying, 2002). When the right cluster method is applied it allows for high performance and scalability which is important when considering expanding beyond a given sample into others while retaining high accuracy levels.

As cluster analysis takes on many forms, it is necessary to ensure that the most accurate technique is being utilised to ascertain that the target subsample, in this case the most impoverished students, has been accurately identified. Therefore, using fuzzy logic to create the baseline and identify a candidate dependent variable set allows for a subsequent comparison of the various cluster analysis techniques considered herein. The goal of data selection is to separate relevant objects from not relevant ones (Hudec and Vujoševic, 2012), which allows for their dichotomisation for further use.

## 3 Establishing baseline group of impoverished students

### *3.1 Survey*

The survey being analysed is conducted periodically in schools which are part of the Jesuit-sponsored not-for-profit organisation dedicated to the education of the 'poorest of the poor' throughout Latin America, Fe y Alegría. The analysis described herein involves surveys conducted in six high schools in two different cities in Bolivia. Two high schools (Luiz Espinal Camps and Fray Vicente Bernedo) are in Potosí, a city in the high Andean plateau and four high schools (Sagrada Familia, Jose Maria Velaz, Gualberto Paredes and Loyola de FyA) are in Sucre, a city in the foothills of the Andes. A total of 732 students enrolled in the fourth and fifth grades in 'la secundaria', corresponding to US high-school sophomore and junior classes in each school completed the survey.

Each survey consists of two identifier questions (name and gender) and 22 additional lifestyle and well-being questions. These questions considered a variety of categories: parents' work and education; home life (number of rooms, number of people in a bedroom, study space); basic necessities (e.g., food, water, electricity); income estimates (number of people in household, number who contribute, student with job); and extras (toys, books, phone, weekend activities). A few questions were yes or no questions (scale yes = 1 and no = 2). Other questions were Likert scale questions with either 4 or 5 choices for the answers (usually ranked lowest to highest).

### *3.2 Data cleaning*

In order to prepare the data for analysis, data cleaning was required. A total of 45 students were removed from the data set. Twenty of these students answered survey questions outside the range of a given question. For instance, when asked do you have electricity (1 = yes, 2 = no) the student would give an answer of 3, 4, or 5. The remaining 25 students who were dropped gave inconsistent answers to sets of two or more question. For example, they might have responded that four or more people slept in their bedroom but had previously answered that they lived with 3 people. Assuming that the students understood what the questions were asking, this would act as evidence that they were not sharing entirely accurate data and were therefore removed.

### *3.3 Establishing a dichotomised dependent variable*

Five questions of 22 (Appendix) aligned directly with the UN definition of absolute poverty (UN, 1995). Since students who lack water and electricity were part of a yes/no question, those who lacked these essential items were immediately identified (dichotomous value = 1). The distributions for the remaining questions (number of rooms in a household, number of meals a day and number of earners who contribute weekly) were also considered. Students who fell in the lowest 5% based on their distributions were identified (dichotomous value = 1) using the fuzzy logic steps as expressed in Algorithm 1. This established the baseline for impoverished students. It is imperative to note that the investigation is interested in a larger proportion of the population than those identified through this dichotomous measure. As mentioned above, without any viable dependent variable there was no way to identify which cluster technique best identified the students in need. This baseline allows for a metric to assess the different analyses.

**Algorithm 1** Fuzzy logic applied to aggregate lowest 5 % of UN definition

| | |
|---|---|
| Input → | Distribution of each question pertaining to definition of UN absolute poverty |
| Output → | Set of students who are described as impoverished |
| 1 | Consider each question that falls into the UN definition $Q_d$, {d = 1, 2, 3} |
| 2 | Create distribution of each and test distribution for normality {QQnorm plot}. |
| 3 | IF (distribution $Q_d$ = Normal) |
| 4 | Value = $\left(V_{.05} * \sigma_{Q_d} + \mu\right)$ → All students who answer $Q_d \leq$ Value designate (Y = 1) |
| 5 | IF (distribution $Q_d \neq$ Normal) |
| 6 | Position = .05(n + 1). All students ≤ position in Sort ($Q_d$) → Value designate (Y = 1) |
| 7 | Else |
| 8 | All undefined values are not classified as impoverished (Y = 0) |

## 4 Analysis and selected clustering methods

### *4.1 Data reduction*

In order to make sure that the correct questions are considered, preparatory analysis is required. Correlation between all pairs of questions, a quantity measuring the extent of association between variables, varying from –1 (no association) to 1 (complete association) where the closer a value is to the absolute value of 1 the stronger the association between the two variables, was considered. All variables with an absolute correlation value greater than 0.2 were considered to have an acceptable level feasible association in this case. Higher values are not expected due the fact that answer values were either from yes/no questions with answers of (1 or 2) or Likert scale questions with answers of (1 to 4) or (1 to 5). The cross-correlations in Figure 1 are expressed with colour coding according to the value of respective correlation.

**Figure 1** Cross-correlation matrix that looks the correlation between every survey question pair (see online version for colours)

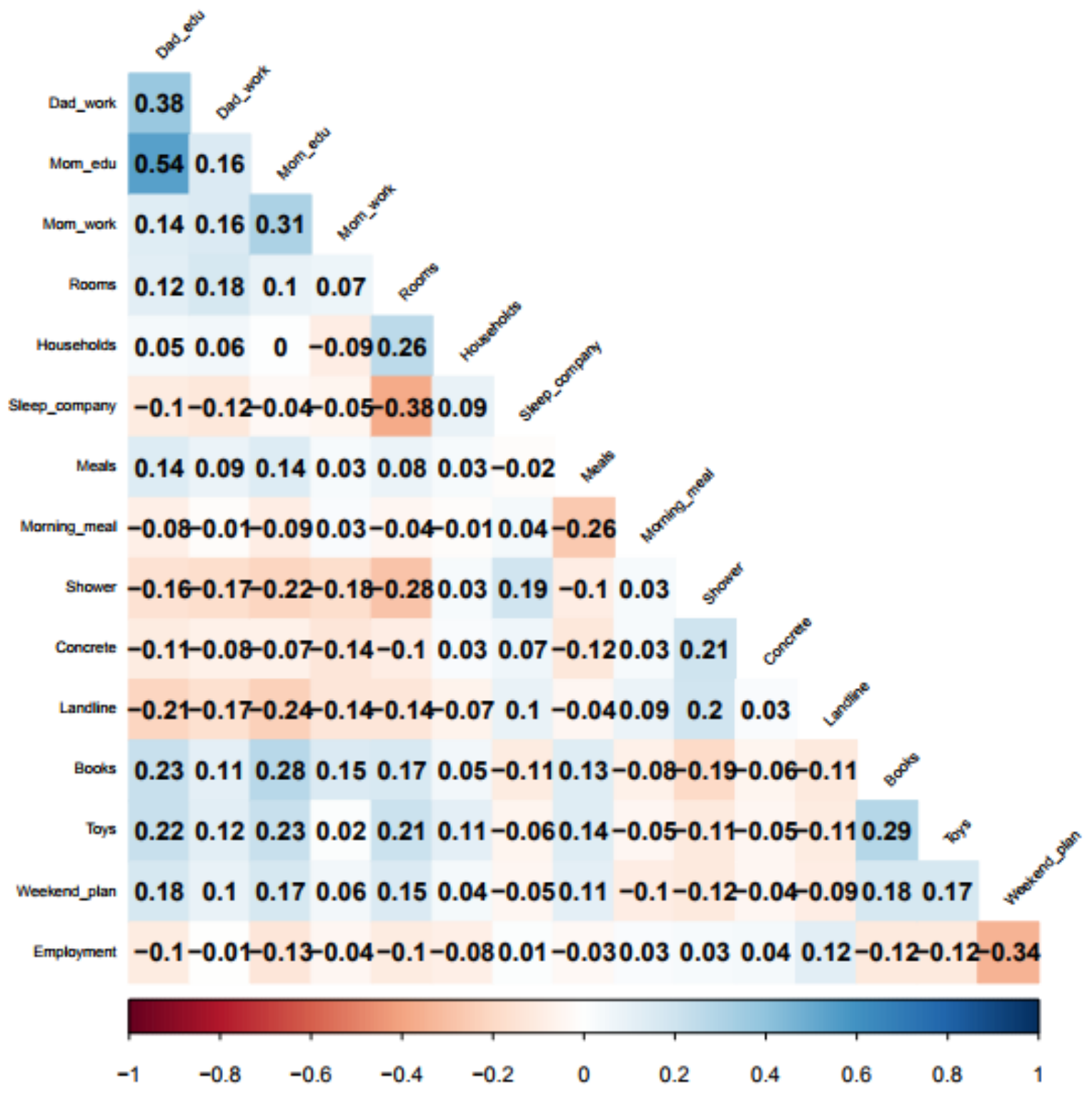

A cross-correlation matrix of all survey questions identified minimal multicollinearity among the questions as depicted in Table 1. The strongest and most likely correlated questions were between the two parents' education (0.54), perhaps an indication that parents tended to marry people within the same educational strata. In order to minimise double-counting the variability from these variables in the remaining analysis, the father's education was removed from the questions.

**Table 1** Correlation table showing the paired correlations among the survey questions

| *Question$_i$* | *Question$_j$* | *Correlation* |
|---|---|---|
| Dad_edu | Mom_edu | 0.540185 |
| Dad_edu | Dad_work | 0.37984 |
| Rooms | Sleep_company | –0.379356 |
| Weekend_plan | Employment | –0.3416168 |
| Mom_edu | Mom_work | 0.3131052 |
| Books | Toys | 0.286035 |
| Rooms | Shower | –0.2820303 |
| Mom_edu | Books | 0.2770261 |
| Rooms | Households | 0.2643342 |
| Meals | Morning_meal | –0.2617946 |
| Mom_edu | Landline | –0.2382998 |
| Mom_edu | Toys | 0.2290978 |
| Dad_edu | Books | 0.2250642 |
| Dad_edu | Toys | 0.2244639 |
| Mom_edu | Shower | –0.2187576 |
| Rooms | Toys | 0.2149864 |
| Dad_edu | Landline | –0.2111559 |
| Shower | Concrete | 0.2064452 |

Factor analysis operates based on the notion that measurable and observable variables can be reduced to fewer latent variables that share a common variance and are unobservable, a reduction in dimensionality (Yong and Pearce, 2013). As in all statistical approaches, EFA is a complex process with few absolute guidelines and many options (Osborne and Costello, 2005). For the purposes of this study, the R program was used and the steps and methods of the EFA are as follows.

First, an initial principal components analysis (PCA) with Varimax rotation was considered. In order to determine the number of factors to retain, the most commonly used criterion considers the number of factors to be equivalent to the number of eigenvalues with values above one, known as the Kaiser criterion (Kaiser, 1960). For this data set, R code (Princomp function) was applied and all 21 principal components were calculated/considered. Eight eigenvalues had numeric values greater than 1 which is considered as significant and led to analysis using eight principal components and hence considering a factor analysis with n = 8 factors.

These eight factors are made orthogonal using the Varimax operator in order to simplify the expression without changing relative values. Next, the variables (survey questions) that provide the most meaningful insights as to the variability among the students are identified within each factor (principal component) using a threshold value of 0.30: any variable (survey question) in a relevant factor (component) with value below .30 was ignored (see Table 2). The questions excluded from further analysis because they provided little information on each of the factors were 'electricity', 'meals', 'desk',

'concrete', 'landline', 'earners' and 'gangs'. See Table 2 for the relative presence of each survey question (factor) in each of the identified principal components.

**Table 2** Factor loadings of the eight orthogonal factors identified in the PCA

| *Loadings* | *Factor 1* | *Factor 2* | *Factor 3* | *Factor 4* | *Factor 5* | *Factor 6* | *Factor 7* |
|---|---|---|---|---|---|---|---|
| Rooms | 0.662 | | | | 0.33 | | |
| Sleep_Company | –0.608 | | | 0.52 | | | |
| Employment | | 0.997 | | | | | |
| Morning_meal | | | 0.993 | | | | |
| Households | | | | | 0.52 | | |
| Dad_work | 0.312 | | | | | | |
| Mom_edu | 0.326 | | | 0.434 | –0.331 | | |
| Mom_work | | | | | –0.326 | | |
| Water | | | | | | 0.337 | |
| Family_members | | | | | | 0.4 | |
| Shower | –0.46 | | | | | | |
| Books | 0.347 | | | | | | |
| Toys | 0.317 | | | | | | |
| Weekend_plan | | –0.345 | | | | | |
| SS loadings | 1.792 | 1.274 | 1.148 | 0.928 | 0.82 | 0.459 | 0.374 |
| Proportion var. | 0.085 | 0.061 | 0.055 | 0.044 | 0.039 | 0.022 | 0.018 |
| Cumulative var. | 0.085 | 0.146 | 0.201 | 0.245 | 0.284 | 0.306 | 0.344 |

At this point the initial 22 questions were reduced to fourteen questions, which are considered in the remaining analysis.

## 4.2 Cluster analysis

Though cluster analysis is very useful for data segmentation, two challenges are selecting the most appropriate technique to apply and identifying the correct number of clusters. For this study three specific cluster techniques are considered: k-means, k-modes and hierarchical clustering (using three different hierarchical variations). The hierarchical methods are complete linkage-based, single linkage-based and mean linkage-based. All five techniques were considered and compared. Recall that this data is unsupervised and effort was put into identifying students who could feasibly be described as in absolute poverty using fuzzy logic. These five-clustering technique will be compared in an effort to ascertain if one provides a better methodology for identifying the aforementioned students.

### 4.2.1 K-means cluster analysis

The k-means application was run using the standard 'stats' package in R. One drawback of using k-means is that the user needs to determine how many clusters should be created. In this case, trials were run starting with clusters of size 4 and increasing to clusters as large as 10. The initial request from Fe y Alegria leading to this study was whether one could find the students in the lowest 25th percentile. Since the k-means partitioning often results in relatively similar sized clusters, the cluster sizes of 4, 5 and 6 are represented and discussed as they created clusters of the approximate size requested. It is important to note that regardless of size up to clusters k = 10, the same group of impoverished students were continually identified.

| K-means steps – Steps applied to produce k-means clusters | |
|---|---|
| Input → | $X_1, X_2, X_3, \ldots, X_n$ are the set points |
| Changing variable → | $V_1, V_2, V_3, \ldots, V_c$ is the set of all feasible centre |
| Output → | K-mean clusters |
| 1 | Choose a value for 'c' representing the number of clusters to be created. |
| 2 | Computer randomly assigns initial starting point for the 'c' centres. (Note: for reproducible results, setting a seed is required). |
| 3 | The distance of each data point to the cluster centres is calculated and data points are assigned to the nearest cluster centre. |
| 4 | Recalculate new cluster centres using: |
| | $V_i = \frac{1}{C_i}\sum_{j=1}^{C_i} X_i$, where $c_i$ represents the number of data points in the $i^{\text{th}}$ cluster. |
| 5 | Using position of new data centres, recalculate and assign each data point (as in step 3). |
| 6 | If no data point is re assigned to a new cluster, stop. Else go to step 4. |

### *4.2.2 K-means cluster results*

Cluster of size four produced a most impoverished group including 142 of the 700 students (20.29%). Looking closely at the cluster output in Table 3, there is evidence that cluster 4 consists of students whose fathers traditionally work less (have less stable job environment) and much smaller homes (more than 2.5 fewer rooms on average than any other cluster), they have less access to water and 50% are without shower facilities, more family members sleeping in a bedroom and also fewer toys in the household. Looking more closely at these 142 students, 61 match the original designation of absolute poverty (See Table 4). Hence, this cluster identified 61 of the 120 (50.8%) students initially dichotomized into the need group.

Almost replicate results were found as the clusters continued to be partitioned. When 5 clusters were created using the k-means algorithm, 134 of the 700 students (19.1%) were in the cluster with the least rooms, food and earners. Meanwhile, 58 of the 120 (48.3%) absolute impoverished students were identified. Similarly, when 6 clusters were considered, 129 of the 700 (18.4%) were in the cluster with least food, rooms and earners which identified 58 of the 120 (48.3%) absolute most impoverished students. Clusters results of size five and six can be found in the Tables A1 to A4 in Appendix.

**Table 3** Cluster partitions using k-means where k = 4 clusters

*K-means clustering with four clusters of size 168, 248, 142, 142*

| *Cluster* | *Dad_work* | *Mom_edu* | *Mom_work* | *Rooms* | *House_holds* | *Water* | *who_Live_with* | *#_Sleep* | *early_meal* | *Bath/shwr* | *Books* | *Toys* | *Wknd_plan* | *Job* |
|---|---|---|---|---|---|---|---|---|---|---|---|---|---|---|
| 1 | 3.119 | 3.396 | 3.417 | 4.625 | 2.970 | 1.012 | 1.387 | 1.458 | 1.196 | 1.065 | 3.333 | 2.607 | 3.226 | 1.958 |
| 2 | 2.728 | 2.188 | 1.357 | 4.698 | 3.456 | 1.040 | 1.294 | 1.540 | 1.202 | 1.315 | 2.883 | 2.532 | 3.230 | 1.661 |
| 3 | 2.806 | 2.141 | 1.676 | 4.486 | 3.155 | 1.049 | 1.451 | 1.556 | 1.239 | 1.296 | 2.352 | 2.324 | 1.824 | 3.486 |
| *4* | *2.415* | *2.204* | *1.873* | *1.915* | *2.423* | *1.092* | *1.606* | *2.225* | *1.268* | *1.5* | *2.528* | *2.085* | *2.514* | *2.430* |

Note: Row cluster 4 is most impoverished.

**Table 4** Count of 'absolute' poverty students that fell into each 'k-means' cluster, where k = 4 clusters

| *Reason for student identification* | *1* | *2* | *3* | *4* | *SUM* |
|---|---|---|---|---|---|
| Living in single room | 0 | 0 | 0 | 36 | 36 |
| None of the family members constantly contributes to the family income | 3 | 10 | 5 | 11 | 29 |
| Only one meal per day | 7 | 11 | 7 | 7 | 32 |
| No electricity | 0 | 1 | 0 | 3 | 4 |
| No access to water | 2 | 8 | 5 | 4 | 19 |
| SUM | 12 | 30 | 17 | 61 | 120 |

### *4.2.3 k-modes cluster analysis*

Using the klaR package (Weihs et al., 2005) in R, the k-modes implementation was run in a similar fashion as to that of k-means, that is, no clear rules exists to assess the number of clusters that would be best, so clusters of size 4 through 10 were considered. Since each cluster size produced similar results, only the cluster of size four will be presented in this description, while clusters of size 5 and 6 can be seen in the Tables A5 to A8 in Appendix. There are arguments supporting the traditional approach that converting categorical data into numeric values does not necessarily produce meaningful results (as would be the case with survey data in which some of the questions have no natural order). Huang's (1997) k-modes algorithm, applied herein, allows for the consideration of this possible limitation while extending the k-means paradigm to this categorical data set.

The k-modes algorithm partitions the objects into k groups such that the distance from objects to the assigned cluster modes is minimised. By default, simple-matching distance is used to determine the dissimilarity of two objects. It is computed by counting the number of mismatches in all variables (Weihs et al., 2005). Alternatively, this distance is weighted by the frequencies of the categories in data (Huang, 1997). If an initial matrix of modes is supplied, it is possible that no object will be closest to one or more modes. In this case, fewer clusters than supplied modes will be returned and a warning is given. This is similar to k-means clustering except for slight variation in the distance formula:

$$d(A, B) = \sum_{j=1}^{m} \delta\left(a_j, b_j\right)$$

where

a $\delta\left(a_j, b_j\right) = \begin{pmatrix} 0 & \left(a_j = b_j\right) \\ 1 & \left(a_j \neq b_j\right) \end{pmatrix}$

b $d(a, b)$ is the distance between elements c $A$ and $B$ are two sets of elements (clusters).

**Table 5** Cluster partitions using k-modes where k = 4 clusters

*K-modes clustering with four clusters of size 262, 85, 170, 183*

| *Cluster* | *Dad_work* | *Mom_edu* | *Mom_work* | *Rooms* | *House_holds* | *Water* | *who_Live_with* | *#_Sleep* | *early_meal* | *Bath/shwr* | *Books* | *Toys* | *Wknd_plan* | *Job* |
|---|---|---|---|---|---|---|---|---|---|---|---|---|---|---|
| 1 | 3 | 2 | 1 | 5 | 3 | 1 | 1 | 1 | 1 | 1 | 2 | 2 | 3 | 2 |
| 2 | 2 | *1* | *1* | *1* | 2 | *1* | *1* | *1* | *1* | *1* | 2 | 2 | *1* | *4* |
| 3 | 2 | 2 | 1 | 5 | 3 | 1 | 1 | 2 | 1 | 2 | 2 | 3 | 3 | 1 |
| 4 | 3 | 3 | 3 | 5 | 2 | 1 | 1 | 1 | 1 | 1 | 4 | 3 | 3 | 1 |

Note: Row 2 appears to be most impoverished.

### *4.2.4 K-modes cluster results*

As can be seen in Table 5, a definitive cluster is not absolutely identifiable as to which cluster has the students with most need. While cluster 2 has a mode of one room per household (four fewer than every other cluster) and least educated mother, it shares the lowest tier educated father along with cluster 3. Cluster 2 has the high expectation of being expected to have a job or work for pay (employment value two higher than all others). Though it is not entirely interpretable as a mode, it appears that cluster 2 is the most impoverished group. This cluster has 85 out of 700 (12%) and only accounts for 26 of the 120 need students (21.67%) of the pre-identified students (see Tables 5 and 6). Though other clusters could be chosen using Table 6 after it is generated to maximise the 'correctly' identified students, the cluster modes do not imply that these are the neediest clusters.

**Table 6** Count of students identified as in need in cluster size k = 4

| *Reason for student identification* | *1* | *2* | *3* | *4* | *SUM* |
|---|---|---|---|---|---|
| Living in single room | 9 | *13* | 9 | 5 | 36 |
| None of the family members constantly contributes to the family income | 12 | *6* | 5 | 6 | 29 |
| Only one meal per day | 8 | *4* | 10 | 10 | 32 |
| No electricity | 0 | *3* | 1 | 0 | 4 |
| No access to water | 5 | *0* | 10 | 4 | 19 |
| SUM | 34 | *26* | 35 | 25 | 120 |

### *4.2.5 Hierarchical Cluster Analysis*

Hierarchical clustering (also called hierarchical cluster analysis or HCA) is a method of cluster analysis which seeks to build a hierarchy of clusters (Thomas and Harode, 2015). The three 'shortest distance' rules considered for this analysis are complete linkage, single linkage and mean linkage. The shortest distance definition is the basis for the differentiation between each of these agglomerative clustering methods. The formula for each method can be found in Table 7.

| Hierarchical cluster steps – Steps applied to produce hierarchical clusters | |
|---|---|
| Given → | A set of N items to be clustered and an N * N distance (or shortest distance) matrix, the basic process of hierarchical clustering (defined by Johnson, 1967) is this |
| 1 | Assign each item (student) to a cluster creating N clusters with one item in each. |
| 2 | Find the closest (shortest distance) pair of clusters and merge them into a single cluster (reducing the number of clusters by one). The similarity formula which is defined as the 'shortest distance' depending on the type of hierarchical clustering defined. |
| 3 | Compute the distances (shortest distance as defined) between the new cluster and each of the old clusters. |
| 4 | Repeat steps 2 and 3 until all items are clustered into 'k' number of clusters using the 'shortest distance' rule defined by the cluster type. |

In complete linkage, also known as farthest neighbour, all element pairs are considered where each point in the first cluster is paired with every point in the second cluster. The distance between the clusters is equivalent to the distance between two elements (one from each cluster) that is the farthest away from one another. The shortest set of points (or clusters) are fused together to form new clusters. The steps are repeated until the predetermined k clusters are identified [see Figure 2(a)].

**Figure 2** Visual showing how the hierarchical linkage methods work

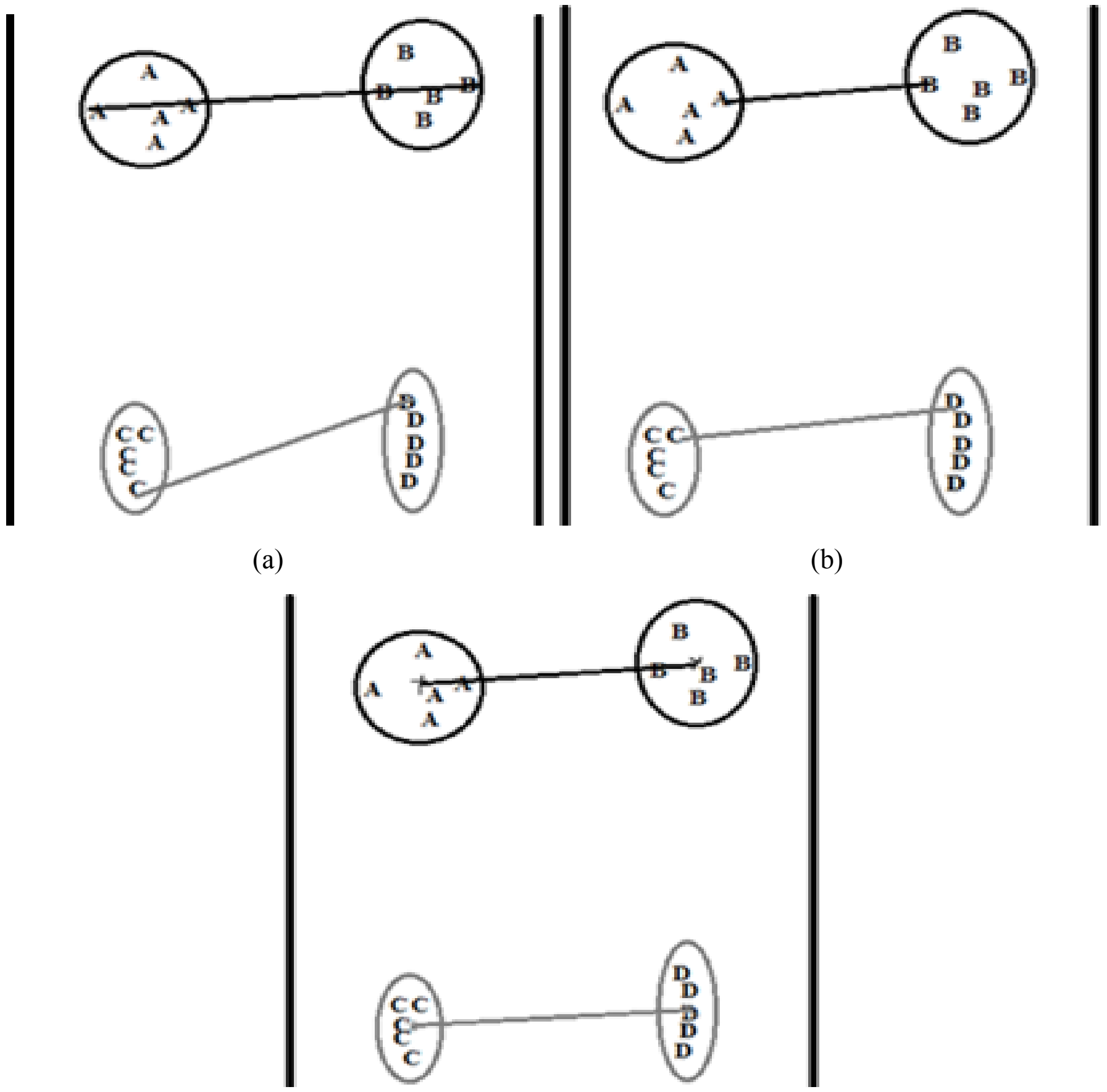

(a) (b)

(c)

Notes: (a) In complete linkage, the cluster with the shortest distance for the farthest neighbor is combined. In the above, cluster of the letter C would be combined with the letter D in the next step, (b) In single linkage, the cluster with the shortest distance from the closest neighbor is combined. In the above, clusters of the letter A would be combined with the letter B in the next step, (c) In mean linkage, the cluster with the shortest distance between cluster average is combined. In the above, clusters of the letter C would be combined with the letter D in the next step.

In single linkage based clustering, also known as the nearest neighbour, all element pairs are again considered where each point in the first cluster is paired with every point in the second cluster. The distance between the clusters is equivalent to the distance between two elements (one from each cluster) that is the nearest to one another. That is that at each step, the two elements or clusters separated by the shortest distance is combined or fused to form new clusters. The steps are repeated until the predetermined k clusters are identified [see Figure 2(b)].

In mean linkage based clustering, also known as average linkage based clustering or the UPGMA (unweighted pair group method with arithmetic mean), all element pairs are again considered where each point in the first cluster is paired with every point in the second cluster. Initially, points closest to one another form into clusters making every paired comparison. The average of each point within each cluster is considered and the two clusters with the smallest difference in averages are not joined or fused. That is that at each step, the two elements or clusters whose mean value of elements is the least different are fused to form new clusters. The steps are repeated until the predetermined k clusters are identified [see Figure 2(c)]. The results for clusters of size 5 and 6 each identified 36 students as well.

**Table 7** Formulas for each hierarchical cluster type considered

| *Complete linkage* | *Single linkage* | *Mean linkage* |
|---|---|---|
| $\max\{d(a,b): a \in A, b \in B\}$ | $\min\{d(a,b): a \in A, b \in B\}$ | $\frac{1}{\lvert A \rvert \lvert B \rvert} \sum_{a \in A} \sum_{b \in B} d(a,b)$ |

Notes: where, d(*a*, *b*) is the distance between elements and *A* and *B* are two sets of elements (clusters).

### *4.2.6 Hierarchical cluster results*

Of the three linkage methods, only the complete linkage gave any form of useable results. That is that the computer returned warning messages and produced fewer clusters than requested. The cluster sizes consisted of one huge cluster and a few tiny clusters (sizes of n = 1, 2, or 3 students). Therefore, only complete linkage is considered in the comparative cluster analysis. For the complete linkage method, cluster sizes of 4, 5 and 6, were again considered. The results were comparable to one another, which can be viewed in Tables 8 and 9 for the cluster of size k = 4 and in the Tables A9 to A10 in Appendix for sizes of k = 5 and 6 respectively. Table 8 provides evidence that Cluster 2 has fewer rooms in their respective households (more than two and half less than every other cluster), less access to water (15% were without access), less likely to have a morning meal, access to a shower and more slightly more likely to work on the weekend. This cluster only identifies 34 of the 120 (28.3%) students initially dichotomised into the need group.

In this section, three linkage methods were considered. The complete linkage provided output data that was somewhat like that seen in both the k-means and the k-modes. However, the single linkage and mean linkage methods provided output that had multiple empty clusters as well as clusters of size one. Therefore, though they provided almost 99% accuracy in predicting the students in the dichotomised need group, the results were exaggerated by the fact that the main cluster contained at minimum 94% of the students (see Appendix). Therefore, these two clustering methods provided unusable results as we would be predicting practically every student as most impoverished while the goal is to identify approximately the lowest 20%.

**Table 8** Cluster partitioning based on complete linkage, where k = 4 clusters

*K-modes clustering with four clusters of size 262, 85, 170, 183*

| *Cluster* | *Dad_work* | *Mom_edu* | *Mom_work* | *Rooms* | *House_holds* | *Water* | *who_Live_with* | *#_Sleep* | *early_meal* | *Bath/shwr* | *Books* | *Toys* | *Wknd_plan* | *Job* |
|---|---|---|---|---|---|---|---|---|---|---|---|---|---|---|
| 1 | 2.61 | 2.58 | 2.36 | 4.32 | 2.8 | 1.03 | 1.48 | 1.59 | 1.23 | 1.27 | 2.84 | 2.41 | 2.38 | 2.4 |
| *2* | *2.15* | *1.89* | *1.69* | *1.43* | *1.65* | *1.15* | *2.15* | *1.91* | *1.3* | *1.58* | *2.43* | *1.76* | *2.13* | *2.89* |
| 3 | 2.67 | 2.13 | 1.4 | 4.20 | 3.43 | 1.05 | 1.28 | 1.68 | 1.19 | 1.33 | 2.65 | 2.42 | 3.17 | 2.04 |
| 4 | 3.67 | 3.16 | 2.63 | 4.19 | 3.52 | 1.02 | 1.2 | 1.69 | 1.2 | 1.09 | 3.24 | 2.66 | 3.32 | 2.09 |

Note: Row cluster 2 is most impoverished.

## 5 Comparisons and conclusions

In comparing the methods, the k-means method gave a higher percentage of correctly identified students from the original poverty dichotomisation. Within each question used to identify this baseline set, it accurately identified the students living in a single room with 100% accuracy which was much higher than the other two methods, 36% and 61.1%% respectively. The k-means method was better at identifying the students with less earners in the household (38%) while k-modes and complete linkage were 21% and 24% respectively. K-means also had a higher proportion of students who had only one meal a day (22%) when compared to k-modes and complete linkage at (12.5% and 3% respectively). With respect to electricity, all methods identified three of the four students. As for water, k-means identified 21%, complete linkage only 3% and k-modes found 0%. Overall, in every specific groups k-means matched or exceeded all other measures when considering this form of proper identification of the students. In total, k-means identified 50.8%, which exceeded the other methods by more than 20%.

**Table 9** Hierarchical cluster partitions using 'complete linkage' where k = 4 clusters

| *Reason for student identification* | *1* | *2* | *3* | *4* | *SUM* |
|---|---|---|---|---|---|
| Living in single room | 1 | 22 | 5 | 8 | 36 |
| None of the family members constantly contributes to the family income | 9 | 7 | 9 | 4 | 29 |
| Only one meal per day | 13 | 1 | 13 | 5 | 32 |
| No electricity | 0 | 3 | 1 | 0 | 4 |
| No access to water | 8 | 1 | 8 | 2 | 19 |
| SUM | 31 | 34 | 36 | 19 | 120 |

**Table 10** Comparison of each useable methods for cluster size k = 4

| *Reason for identification of student* | *k-means* | *k-modes* | *Complete linkage* | *Max findable* |
|---|---|---|---|---|
| Living in single room | 36 (100%) | 13 (36%) | 22 (61.1%) | 36 |
| No family members constantly contribute to family income | 11 (38%) | 6 (21%) | 7 (24%) | 29 |
| Only one meal per day | 7 (22%) | 4 (12.5%) | 1 (3%) | 32 |
| No electricity | 3 (75%) | 3 (75%) | 3 (75%) | 4 |
| No access to water | 4( 21%) | 0( 0%) | 1 (5.3%) | 19 |
| Sum | 61 (50.8%) | 26 (21.7%) | 34 (28.3%) | 120 |

In Figure 3, the lines represent the percentage of the original dichotomised students that were captured by each cluster type. While complete linkage (red) and k-modes provided somewhat similar results of approximately 20 to 30% of students correctly identified, k-means provided a continuously higher percentage of about 50% of students correctly identified despite changing number of clusters produced. Overall, this method provided more consistent results.

**Figure 3** Percentage of correctly identified students for cluster sizes of k = 4, 5 and 6 (see online version for colours)

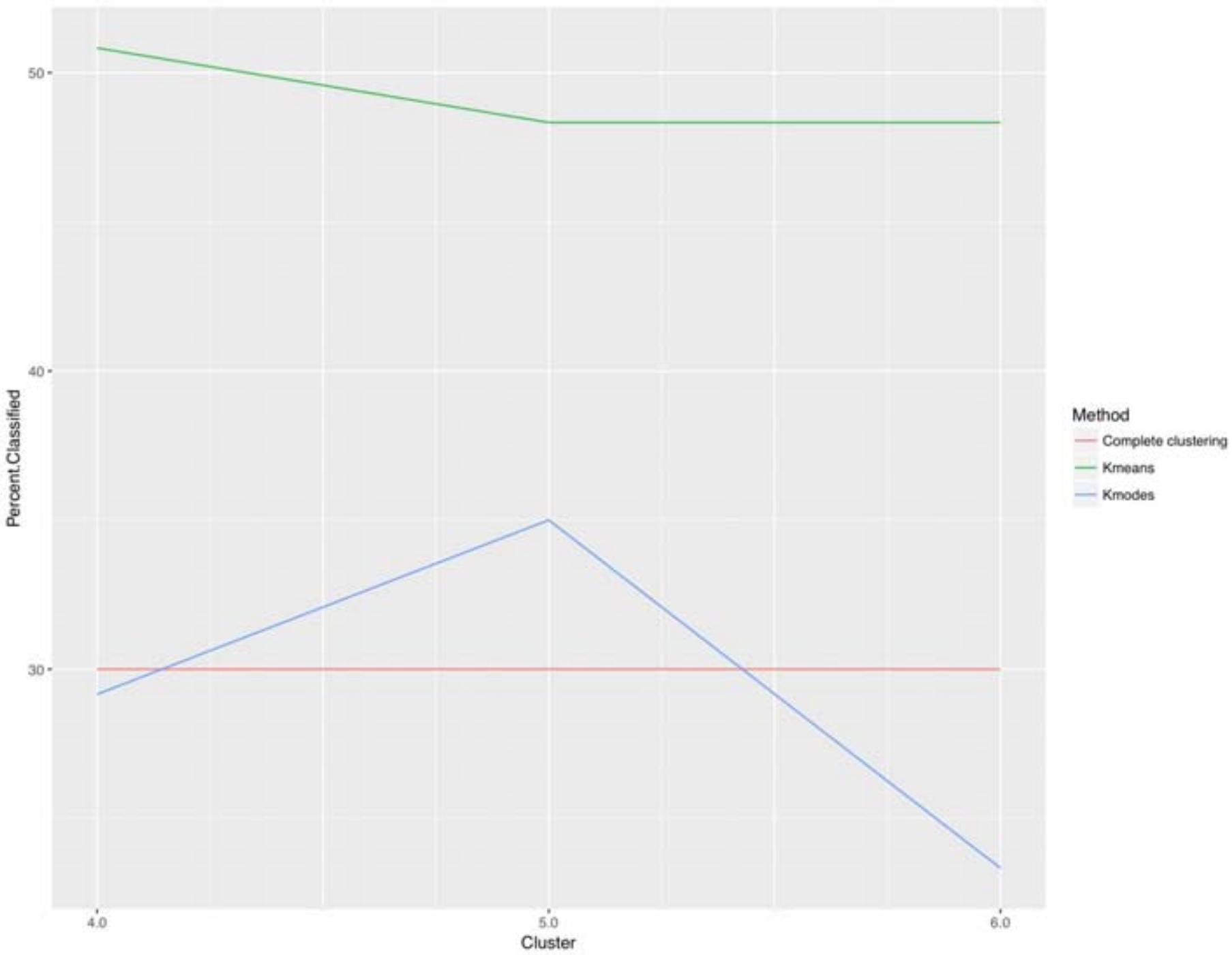

All of the methods mentioned herein provided insightful though not completely accurate results and the k-means method provided significantly better results. The resultant students from both the fuzzy logic group and the k-means group were shared with Fe y Alegría, the group who had requested identification which initiated this analysis. However, efforts continue to make a more cohesive collection allowing for a baseline to be created each time the survey is given and an accurate cluster analysis technique to be applied to help identify both the absolute poverty students and those students with comparable though less extreme situations. At this time, k-means has been consistently applied to each sample analysed.

## 6 Issues

While the survey was created in an effort to collect varying information about the students, the initial survey intention was not to identify poverty as it was a subset of a broader survey. With more refined questions, a better solution may be feasible. For example, access to water means different things to different people. Walking half a mile to a stream to collect daily household water may be considered access to some students while other students may have a faucet within the vicinity close to the dwelling. Being more specific about the distance and access may be able to put this question on a five-point scale keeping it relevant in the study. Moreover, availability of a cell phone and how many times it is used per week as a question may garner more insight than whether there is access to a landline. Finally, question one asks for the father's education level and question two asks the work status. Both questions assume the father is alive and part of the family make-up. However, a decent number of students skipped this combination (and/or the mother combination) leading the analysts to question whether these students might be in a single family home which would be insightful in a third world country when considering poverty. Overall, the survey is currently being restructured for future application and analysis.

Another small concern is the validity of answers from students who may not wish to share information or may not have the survey language as a first language, leading to potential misunderstandings. Some students would answer invalid responses – like a 3, 4, or 5 in answer to a yes/no question. It is hard to tell why these students didn't answer correctly so for this study they have been omitted. Future survey applications will ensure that students answer all questions and that attention is paid when responses are given in order to identify mislabelled answers more quickly for analysis.

## 7 Future research

Continued effort and analysis are planned to ascertain a better overall cluster technique to identify the most impoverished. Once the baseline hits a higher threshold of correct responses (greater than 90%), bootstrapping techniques will be applied to sampled groups within the larger data set where clusters will be created and ranked from least to greatest. These rankings will be recorded and the bootstrap will be run thousands of times with the goal of creating a more continuous (over dichotomous) measure for each student as an established dependent variable. Ideally the results will allow for regression models to help identify the significance and effect of each imperative question with the goal of making a model to help identify the most impoverished students.

## References

AL-Sharuee M., Liu, F. and Pratama, M. (2018) 'An automatic contextual analysis and ensemble clustering approach and comparison', *Data and Knowledge Engineering*, May, Vol. 115, pp.194-213 [online] https://doi.org/10.1016/j.datak.2018.04.001 (accessed 7 April 2018).

Huang, Z. (1997) 'A fast clustering algorithm to cluster very large categorical data sets in data mining', in Lu, H., Motoda, H. and Luu, H. (Eds.): *KDD: Techniques and Applications*, pp.21–34, World Scientific, Singapore.

Hudec, M. and Vujoševic´, M. (2012) 'Expert systems with applications', *Expert Systems with Applications*, Vol. 39, No. 10, pp.8817–8823.

Johnson, S.C. (1967) 'Hierarchical clustering schemes', *Psychometrika*, Vol. 32, No. 3, pp.241–254.

Kaiser, H.F. (1960) 'The application of electronic computers to factor analysis', *Educational and Psychology Measurement*, Vol. 20, No. 1, pp.141–151.

Lovett S., Zeiss A. and Heinemann G. (2002) 'Assessment and development: now and in the future', in Zeiss, A.M. and Heinemann G.D. (Eds.): *Team Performance in Health Care, Issues in the Practice of Psychology*, Springer, Boston, MA.

Lughofer, E., Pratama, M. and Skrjanc, I. (2017) 'Incremental rule splitting in generalized evolving fuzzy systems for autonomous drift compensation', *IEEE Transactions on Fuzzy Systems*, Vol. 26, No. 4.

Major, M. and Sedlacek, W. (2001) 'Using factor analysis to organise students' services', *Journal of College Student Development*, Vol. 42, No. 3, pp.2272–2278.

Morris, S. (2001) 'Sample size required for adverse impact analysis', *Applied HRM Research*, Vol. 6, No. 1, pp.13–32.

United Nation (UN, 1995) *The Copenhagen Declaration and Programme of Action: World Summit for Social Development*, United Nations Department of Publications, New York.

Osborne, J.W. and Costello, A.B. (2005) 'Best practices in exploratory factor analysis: four recommendations for getting the most from your analysis', *Pan-Pacific Management Review*, Vol. 12, No. 12, pp.131–146.

Sharma, N., Bajpai, A. and Litoriya, R. (2012) 'Comparing the various clustering algorithms of weka tools', *International Journal of Emerging Technology and Advanced Engineering*, Vol. 2, No. 5, pp.73–80.

Thomas, S.T. and Harode, U. (2015) 'A comparative study in K-means and hierarchical clustering', *International Journal of Electronics and Computational Systems*, Vol. 4, No. 2, pp.5–11.

Vyas, S. and Kumaranayake, L. (2006) 'Constructing socio-economic status indices: how to use principal component analysis', *Health Policy Plan*, Vol. 6, No. 11, pp.459–468.

Weihs, C., Ligges. U., Luebke, K. and Raabe, N. (2005) 'klaR analysing german business cycles', in Baier, D., Decker, R. and Schmidt-Thieme, L. (Eds): *Data Analysis and Decision Support*, pp.335–343, Springer-Verlag, Berlin.

Wei-ning, Q. and Ao-ying, Z. (2002) 'Analysing popular clustering algorithms from different viewpoints', *Journal of Software*, Vol. 13, No. 8, pp.1382–1393.

Yong, A. and Pearce, S. (2013) 'A beginner's guide to factor analysis: focusing on exploratory factor analysis', *Tutorials in Quantitative Methods for Psychology*, Vol. 9, No. 2, pp.79–94.

## Appendix

**Table A1** Cluster partitions using 'k-means' where k = 5 clusters

| *K-modes clustering with four clusters of size 262, 85, 170, 183* | | | | | | | | | | | | | | |
|---|---|---|---|---|---|---|---|---|---|---|---|---|---|---|
| *Cluster* | *Dad_work* | *Mom_edu* | *Mom_work* | *Rooms* | *House_holds* | *Water* | *who_Live_with* | *#_Sleep* | *early_meal* | *Bath/shwr* | *Books* | *Toys* | *Wknd_plan* | *Job* |
| 1 | 3.12 | 3.6 | 3.49 | 4.62 | 2.87 | 1 | 1.38 | 1.46 | 1.21 | 1.06 | 3.41 | 2.61 | 3.17 | 1.84 |
| 2 | 2.74 | 2.28 | 1.55 | 4.72 | 4.58 | 1.04 | 1.24 | 1.67 | 1.18 | 1.36 | 2.97 | 2.52 | 3.17 | 1.78 |
| *3* | *2.48* | *2.22* | *1.89* | *1.87* | *2.4* | *1.08* | *1.62* | *2.27* | *1.26* | *1.5* | *2.54* | *2.06* | *2.5* | *2.43* |
| 4 | 2.75 | 2.17 | 1.75 | 4.39 | 3.22 | 1.05 | 1.51 | 1.56 | 1.26 | 1.29 | 2.43 | 2.42 | 1.5 | 3.64 |
| 5 | 2.82 | 2.09 | 1.48 | 4.57 | 2.46 | 1.04 | 1.33 | 1.41 | 1.19 | 1.24 | 2.65 | 2.44 | 3.26 | 1.92 |

Note: Row cluster 3 is the most impoverished.

**Table A2** Count of 'absolute' poverty students that fell into each 'k-means' cluster, where k = 5 clusters

| *Reason student's are identified* | *1* | *2* | *3* | *4* | *5* | *SUM* |
|---|---|---|---|---|---|---|
| Living in single room | 0 | 0 | 36 | 0 | 0 | 36 |
| None of the family members constantly contributes to the family income | 2 | 4 | 9 | 5 | 9 | 29 |
| Only one meal per day | 5 | 8 | 7 | 6 | 6 | 32 |
| No electricity | 0 | 1 | 3 | 0 | 0 | 4 |
| No access to water | 0 | 4 | 3 | 4 | 8 | 19 |
| SUM | 7 | 17 | 58 | 15 | 23 | 120 |

**Table A3** Cluster partitions using 'k-means' where k = 6 clusters

*K-means clustering with six clusters of size 166, 101, 128, 129, 95, 81*

| *Cluster* | *Dad_work* | *Mom_edu* | *Mom_work* | *Rooms* | *House_holds* | *Water* | *who_Live_with* | *#_Sleep* | *early_meal* | *Bath/shwr* | *Books* | *Toys* | *Wknd_plan* | *Job* |
|---|---|---|---|---|---|---|---|---|---|---|---|---|---|---|
| 1 | 2.87 | 2.17 | 1.31 | 4.6 | 2.48 | 1.04 | 1.3 | 1.43 | 1.21 | 1.22 | 2.73 | 2.48 | 3.18 | 2.01 |
| 2 | 3.3 | 2.32 | 3.69 | 4.66 | 2.86 | 1 | 1.33 | 1.42 | 1.28 | 1.16 | 2.75 | 2.32 | 2.93 | 2.07 |
| 3 | 2.67 | 2.27 | 1.45 | 4.7 | 4.58 | 1.03 | 1.26 | 1.67 | 1.17 | 1.35 | 3 | 2.53 | 3.2 | 1.77 |
| *4* | 2.39 | *2.16* | *1.85* | *1.85* | *2.42* | *1.1* | *1.6* | *2.26* | *1.26* | *1.51* | *2.51* | *2.04* | *2.48* | *2.44* |
| 5 | 2.63 | 2.11 | 1.61 | 4.33 | 3.38 | 1.05 | 1.56 | 1.61 | 1.27 | 1.29 | 2.37 | 2.41 | 1.36 | 3.72 |
| 6 | 2.81 | 4.48 | 3.02 | 4.48 | 2.77 | 1.01 | 1.45 | 1.5 | 1.09 | 1.08 | 3.72 | 2.8 | 3.39 | 1.72 |

Note: Row cluster 4 is the most impoverished.

**Table A4** Count of ‘absolute’ poverty students that fell into each ‘k-means’ cluster, where k = 6 clusters

| *Reason for student identification* | *1* | *2* | *3* | *4* | *5* | *6* | *SUM* |
|---|---|---|---|---|---|---|---|
| Living in single room | 0 | 0 | 0 | 35 | 0 | 1 | 36 |
| None of the family members constantly contributes to the family income | 10 | 3 | 4 | 9 | 3 | 0 | 29 |
| Only one meal per day | 6 | 3 | 8 | 7 | 4 | 4 | 32 |
| No electricity | 0 | 0 | 1 | 3 | 0 | 0 | 4 |
| No access to water | 7 | 1 | 3 | 4 | 4 | 0 | 19 |
| SUM | 23 | 7 | 16 | 58 | 11 | 5 | 120 |

**Table A5** Cluster partitions using 'k-modes' where k = 5 clusters

*K-modes clustering with 5 clusters of size 331, 114, 112, 101, 42*

| *Cluster* | *Dad_work* | *Mom_edu* | *Mom_work* | *Rooms* | *House_holds* | *Water* | *who_Live_with* | *#_Sleep* | *early_meal* | *Bath/shwr* | *Books* | *Toys* | *Wknd_plan* | *Job* |
|---|---|---|---|---|---|---|---|---|---|---|---|---|---|---|
| 1 | 2 | 2 | 1 | 5 | 3 | 1 | 1 | 1 | 1 | 1 | 4 | 3 | 3 | 1 |
| 2 | 3 | 2 | *1* | 5 | 2 | *1* | *1* | *1* | 2 | *1* | 3 | 2 | 3 | 2 |
| 3 | 2 | 2 | 1 | 3 | 2 | 1 | 1 | 2 | 1 | 2 | 2 | 2 | 1 | 4 |
| 4 | 3 | 2 | 2 | 4 | 2 | 1 | 1 | 2 | 1 | 1 | 2 | 3 | 2 | 2 |
| 5 | 2 | 3 | 3 | 5 | 2 | 1 | 2 | 2 | 1 | 1 | 2 | 2 | 3 | 1 |

Note: Row cluster 2 is the most impoverished

**Table A6** Count of 'absolute' poverty students that fell into each 'k-modes' cluster, where k = 5 clusters

| *Reason for student identification* | *1* | *2* | *3* | *4* | *SUM* |
|---|---|---|---|---|---|
| Living in single room | 9 | 13 | 9 | 5 | 36 |
| None of the family members constantly contributes to the family income | 12 | 6 | 5 | 6 | 29 |
| Only one meal per day | 8 | 4 | 10 | 10 | 32 |
| No electricity | 0 | 3 | 1 | 0 | 4 |
| No access to water | 5 | 0 | 10 | 4 | 19 |
| SUM | 34 | 26 | 35 | 25 | 120 |

**Table A7** Cluster partitions using ‘k-modes’ where k = 6 clusters

| *K-modes clustering with six clusters of size 58, 185, 212, 100, 71, 74* | | | | | | | | | | | | | | |
|---|---|---|---|---|---|---|---|---|---|---|---|---|---|---|
| *Cluster* | *Dad_work* | *Mom_edu* | *Mom_work* | *Rooms* | *House_holds* | *Water* | *who_Live_with* | *#_Sleep* | *early_meal* | *Bath/shwr* | *Books* | *Toys* | *Wknd_plan* | *Job* |
| *1* | *2* | *1* | *1* | *1* | *1* | *1* | *2* | *1* | *1* | *2* | *2* | *2* | *1* | *4* |
| 2 | 2 | 2 | 1 | 5 | 3 | 1 | 1 | 2 | 1 | 1 | 2 | 2 | 3 | 2 |
| 3 | 3 | 2 | 3 | 5 | 3 | 1 | 1 | 1 | 1 | 1 | 4 | 3 | 3 | 1 |
| 4 | 2 | 2 | 2 | 5 | 2 | 1 | 1 | 1 | 1 | 1 | 2 | 3 | 3 | 1 |
| 5 | 2 | 2 | 1 | 5 | 3 | 1 | 1 | 2 | 1 | 2 | 2 | 3 | 2 | 1 |
| 6 | 3 | 2 | 1 | 5 | 5 | 1 | 1 | 1 | 1 | 1 | 3 | 2 | 3 | 1 |

Note: Row cluster 1 is the most impoverished.

**Table A8** Count of 'absolute' poverty students that fell into each 'k-modes' cluster, where k = 6 clusters

| *Reason for student identification* | *1* | *2* | *3* | *4* | *5* | *6* | *SUM* |
|---|---|---|---|---|---|---|---|
| Living in single room | 13 | 11 | 5 | 3 | 2 | 2 | 36 |
| None of the family members constantly contributes to the family income | 8 | 6 | 7 | 4 | 3 | 1 | 29 |
| Only one meal per day | 5 | 5 | 8 | 5 | 8 | 1 | 32 |
| No electricity | 2 | 0 | 0 | 0 | 1 | 1 | 4 |
| No access to water | 0 | 4 | 2 | 5 | 6 | 2 | 19 |
| SUM | 28 | 26 | 22 | 17 | 20 | 7 | 120 |

**Table A9** Cluster partitions using 'hierarchical' where k = 5 clusters

| *Cluster* | *Dad_work* | *Mom_edu* | *Mom_work* | *Rooms* | *Households* | *Water* | *who_Live with* | *#_Sleep* | *early_meal* | *Bath/shwr* | *Books* | *Toys* | *Wknd_plan* | *Job* |
|---|---|---|---|---|---|---|---|---|---|---|---|---|---|---|
| 1 | 2.52 | 2.13 | 1.65 | 4.07 | 3.34 | 1.03 | 1.46 | 1.6 | 1.25 | 1.39 | 2.46 | 2.27 | 1.43 | 3.3 |
| 2 | *2.15* | *1.89* | *1.69* | *1.43* | *1.65* | *1.15* | *2.15* | *1.91* | *1.3* | *1.58* | *2.43* | *1.76* | *2.13* | *2.89* |
| 3 | 2.67 | 2.13 | 1.4 | 4.2 | 3.43 | 1.05 | 1.28 | 1.68 | 1.19 | 1.33 | 2.65 | 2.42 | 3.17 | 2.04 |
| 4 | 2.67 | 2.86 | 2.8 | 4.48 | 2.46 | 1.02 | 1.49 | 1.58 | 1.22 | 1.2 | 3.07 | 2.5 | 2.96 | 1.84 |
| 5 | 3.67 | 3.16 | 2.63 | 4.19 | 3.52 | 1.02 | 1.2 | 1.69 | 1.2 | 1.09 | 3.24 | 2.66 | 3.32 | 2.09 |

Note: Row cluster 2 is the most impoverished.

**Table A10** Count of 'absolute' poverty students that fell into each 'hierarchical' cluster, where k = 5 clusters

| *Reason for student identification* | *1* | *2* | *3* | *4* | *5* | *SUM* |
|---|---|---|---|---|---|---|
| Living in single room | 0 | 22 | 5 | 1 | 8 | 36 |
| None of the family members constantly contributes to the family income | 5 | 7 | 9 | 4 | 4 | 29 |
| Only one meal per day | 9 | 1 | 13 | 4 | 5 | 32 |
| No electricity | 0 | 3 | 1 | 0 | 0 | 4 |
| No access to water | 3 | 1 | 8 | 5 | 2 | 19 |
| SUM | 17 | 34 | 36 | 14 | 19 | 120 |

**Table A11** Cluster partitions using ‘hierarchical’ where k = 6 clusters

| *Cluster* | *Dad_work* | *Mom_edu* | *Mom_work* | *Rooms* | *House_holds* | *Water* | *who_Live with* | *#_Sleep* | *early_meal* | *Bath/shwr* | *Books* | *Toys* | *Wknd_plan* | *Job* |
|---|---|---|---|---|---|---|---|---|---|---|---|---|---|---|
| 1 | 2.52 | 2.13 | 1.65 | 4.07 | 3.34 | 1.03 | 1.46 | 1.6 | 1.25 | 1.39 | 2.46 | 2.27 | 1.43 | 3.3 |
| **2** | *2.15* | *1.89* | *1.69* | *1.43* | *1.65* | *1.15* | *2.15* | *1.91* | *1.3* | *1.58* | *2.43* | *1.76* | *2.13* | *2.89* |
| 3 | 2.67 | 2.13 | 1.4 | 4.2 | 3.43 | 1.05 | 1.28 | 1.68 | 1.19 | 1.33 | 2.65 | 2.42 | 3.17 | 2.04 |
| 4 | 2.67 | 2.86 | 2.8 | 4.48 | 2.46 | 1.02 | 1.49 | 1.58 | 1.22 | 1.2 | 3.07 | 2.5 | 2.96 | 1.84 |
| 5 | 3.76 | 3.24 | 2.65 | 4.62 | 3.55 | 1.02 | 1.19 | 1.32 | 1.15 | 1.04 | 3.34 | 2.75 | 3.36 | 2.12 |
| 6 | 3.17 | 2.7 | 2.47 | 1.82 | 3.35 | 1.05 | 1.29 | 3.7 | 1.47 | 1.41 | 2.7 | 2.17 | 3.11 | 1.88 |

Note: Row cluster 2 is the most impoverished.

**Table A12** Count of ‘absolute’ poverty students that fell into each ‘hierarchical’ cluster, where k = 6 clusters

| *Reason for student identification* | *1* | *2* | *3* | *4* | *5* | *6* | *SUM* |
|---|---|---|---|---|---|---|---|
| Living in single room | 0 | 22 | 5 | 1 | 1 | 7 | 36 |
| None of the family members constantly contributes to the family income | 5 | 7 | 9 | 4 | 3 | 1 | 29 |
| Only one meal per day | 9 | 1 | 13 | 4 | 5 | 0 | 32 |
| No electricity | 0 | 3 | 1 | 0 | 0 | 0 | 4 |
| No access to water | 3 | 1 | 8 | 5 | 2 | 0 | 19 |
| SUM | 17 | 34 | 36 | 14 | 11 | 8 | 120 |